\documentstyle[prl,aps,psfig,multicol]{revtex}
\def\E{{\hbox{I\kern-.2em\hbox{E}}}}

\def\be {\begin{equation}}
\def\ee {\end{equation}}

\begin{document}
\twocolumn[\hsize\textwidth\columnwidth\hsize\csname 
@twocolumnfalse\endcsname
\title{A Stochastic Description for Extremal Dynamics}
\author{Anne Tanguy $^{(1)}$, Supriya Krishnamurthy $^{(2)}$, Patrice
  Abry $^{(3)}$ and St\'ephane Roux $^{(4)}$}  
\vspace{2mm} 

\address{
{\small (1) \em CNRS UMR 5586, Universit\'e Lyon I,  43, 
bvd du 11 Novembre 1918, \\ 69 622 VILLEURBANNE, France} \\
{\small (2) \em LPMMH,  Ecole Sup\'erieure de Physique et de Chimie
  Industrielles, \\  10 Rue Vauquelin, 75231 Paris, France} \\
{\small (3)  \em CNRS UMR 5672, ENS Lyon, 46, all\'ee d'Italie,\\ 
69 364 LYON Cedex 07, France}\\ 
{\small (4) \em  Laboratoire Surface du Verre et Interfaces, UMR
CNRS/Saint-Gobain, \\  39, Quai Lucien Lefranc, 
F-93303 Aubervilliers Cedex, France} \\
}
\maketitle

\begin{abstract}
We show that 
extremal dynamics 
is very well modelled by the ``Linear
Fractional Stable Motion'' (LFSM), a stochastic process 
entirely defined by two exponents that take into account spatio-temporal
correlations in the distribution of active sites. We
demonstrate this numerically and analytically using well-known
properties of the LFSM.
Further, we use this
correspondence to write an exact expression for an n-point correlation
function as well as an equation of fractional order for
interface growth in extremal dynamics.

\end{abstract} 
\vskip2pc]



\noindent  Since the eighties, several
models have been proposed\cite{Schmitt,Ertas,Stefano,Larkin,Hansen,Tang} to
describe in a common language diverse physical situations such as
roughening of a crack front in fracture\cite{Schmitt}, 
wetting front motion on heterogeneous sufaces\cite{Ertas}, 
dynamics of a ferromagnetic
domain wall driven by an external magnetic field \cite{Stefano}, 
motion of vortices
in type-II superconductors\cite{Larkin}, fluid invasion in porous
media\cite{Hansen}, solid
friction\cite{Tang} or even biological evolution\cite{Maslov}. 
All these models propose to explain the
dissipative behaviour of the system by the competition between an
elastic restoring force and a non linear, randomly distributed, 
time-independent, 
pinning force.  In the case of the spreading of a partially wetting
liquid for example, the {\it pinning} force is due to surface chemical
heterogeneities or roughness, and the elastic restoring force 
is a result of the surface
tension at the liquid/vapor interface.  In the case of {\it strong}
pinning it is well known\cite{Joanny} that, when subjected to an 
external driving, the wetting front displays
local instabilities that force
it to advance quasistatically. This property makes it difficult to
handle the problem in the continuum.
In fact, in the
stationary regime, the main contribution to the global displacement is
from jumps of local parts of the chain resulting from these
instabilities\cite{Tang}. 
Recently, Tanguy et al. \cite{Tang}
have proposed to describe this sort of evolution by an {\it extremal} model:
only the site closest to its instability threshold advances.  After a
jump, the instability thresholds of all the sites are modified by the
(elastic) couplings between sites. 
More precisely, in their model (hereafter LREM for {\it Long Range
Elastic Model}), the interface of size $L$ is defined on a discrete
lattice $(x,h)$.  Initially the front $h(x)=0$, and the pinning
forces $f_p(x,t=0)=f_0(x,h=0)$ are assigned 
independently from a flat distribution. 
The site $x_0$ subjected to the {\it minimum} pinning force
(and hence closest to its instability threshold) advances
first, thus $h(x_0) \to h(x_0)+\Delta h$. At this new position,
a new random pinning force is encountered $f_p(x_0,t+\delta
t)=f_0(x_0,h(x_0)+\Delta h)$. The external
loading $F$ on the system, and interactions along the front, produce a
local driving force on each site $x$ proportional to $f(x,t)=F.\int G(x-y)
h(y,t)\,dy$ where the kernel $G(x) \propto |x|^{-b-1}$ accounts for long range
interactions mediated by the medium.
The loading $F$ is then adjusted so that only one site depins
$f(x,t)=f_p(x,t)$; the others remain trapped since 
$f(y,t)\le f_p(y,t)$ for $y\neq x$.  The  dynamics of advancing
the minimum site and readjusting the others is continued indefinitely. 

A wide class of extremal models have
already been studied extensively by Paczuski et al.\cite{Maslov}.
These models include the Bak-Sneppen evolution model \cite{BS} and 
the Sneppen Interface Model \cite{SN}. All these models try to explain 
driven motion under strong pinning by means of a discrete, 
deterministic dynamics. Only one site is active at every instant of time. 
However the ``time'' is only a way to 
index the sequence of events.  Further the extremal condition
can be thought of as a way to retain the information 
of the spatially quenched 
heterogeneities that determine the evolution of the front.
All the information in this sort of dynamics is clearly contained in
the ``activity'' map: a space-time plot of where the front is active
at every instant of time. 

Previous studies regarding
extremal models \cite{Tang,TL}
have shown that most of the relevant information is
contained in the probability density function (hereafter, pdf) of the
activity map.
In the stationary regime, assuming that the activity was
located at $x_0$ at time $t_0$, the probability that it is at $x$ at time
$t$ is:

\be
\label{eq:furub}
p(|x-x_0|,t-t_0)= (t-t_0)^{-1/z}\phi\left(\frac{|x-x_0|}{(t-t_0)^{1/z}}
\right)
\ee
with

\be
 \phi(r)\propto\cases{ r^{-\alpha-1} &for\quad $r\gg 1$\cr
                       r^0 &for\quad $r\ll 1. $\cr}
\end{equation}

While the exponent $\alpha$ controls the asymptotic behavior of 
the time independant function $\phi$, 
$z$ controls the propagation of the activity along the system as a 
function of time and is therefore termed the
``dynamical exponent". The above distribution is self-affine 
and therefore its temporal evolution is
completely defined through 
the exponents $z$ and $\alpha$ \cite{Furuberg}.
Moreover, as can be seen from  Eq. (\ref{eq:furub}), the 
form of the distribution 
of $|x(t)-x_{0}|$  is independant of $t$ for large enough $t$:
it is ``$\alpha$-stable''\cite{Taqqu}. It has been shown in
\cite{Tang} that $\alpha = b$. 
It is hence also possible to consider the
activity as performing a 
``Brownian like''  motion similar in
spirit to studies of anomalous diffusion \cite{bouchaud}.
The relevant parameters here are: the exponent $\alpha$
appearing in the stationary distribution of the distance between 
{\it successive} active sites (distribution of
increments):
\be
\label{eq:fur}
p\left((x(t+1)-x(t))=l \right)\propto l^{-\alpha-1}\,\,,\forall t
\ee
and the exponent $H$, characteristic of the moments:
$$ \langle \vert x(t)-x(t')\vert^a \rangle ^{1/a}\approx \vert t-t'\vert^{H}\,
\,\,{\rm for}\,\,a<\alpha.$$
$H$ accounts for possible temporal statistical 
dependence between jumps.
When there is no temporal correlation between jumps, this distribution is 
Brownian for $\alpha>2$ and a ``Levy flight'' for $0< \alpha\leq 2$.
In the former case $H= 1/2$ and in the latter case it is easy to show,
for example from the asymptotic expression of $p(x,t)$\cite{bouchaud}, 
that $H=1/\alpha$. However when there are temporal
correlations, as in some of the models studied in
\cite{Maslov}, or in the case when elastic interactions are long
ranged \cite{Tang}, this is no longer true and
$H\neq 1/\alpha$. The value of $H$ is hence indicative of the
presence or absence of temporal correlations in the jumps of the
activity. Attempts have been made in the past to understand
the space-time plot of the activity in extremal models as
an uncorrelated Levy flight \cite{Tang,Supriya},
by keeping only the exponent $\alpha$ and  thus assuming
$H=1/\alpha$. 
However, numerically, this hypothesis leads to erroneous quantitative
predictions. Thus there are long range time correlations which have to
be incorporated in the description.
In this Letter,
we introduce and study such a model.

\noindent We define here a {\it Linear Fractional Stable Motion}.
Let $x$ denote a process generated in the following manner:
\begin{equation}
\label{eq:lf}
  x(t) = \sum f(t,u) \eta(u) \, ,  
\end{equation}  
where $\eta(u)$ is an uncorrelated noise with a symmetric distribution 
$p(\eta=x) \sim |x|^{-\alpha-1}$. Since we would like to consider
stationary processes (if our eventual aim is to describe steady
states) we assume that  $f(t,u)=f(t-u)$.
The definition above implies basically that $x$ consists of a sum of
uncorrelated levy jumps weighted in time by $f(t-u)$. 
This weight function therefore controls
the time dependence of the statistical properties of
$x$. 
It is easy to show that the sum in Eq. \ref{eq:lf} can be performed
much as for independant levy flights and
the random variable $X=x(t)$ (given that $X=0$ at $t=0$) 
has a probability density function
\begin{equation}
\label{eq:pLFSM}
p(X,t) = \int \exp(\imath kX) \,  \exp(-\sigma^{\alpha}_t |k|^{\alpha}) dk/2\pi \, . 
\end{equation}
where $ (\sigma_{t})^\alpha = \sum^{t}_{u=0} |f(t-u)|^{\alpha} $.
For the so-called LFSM \cite{Taqqu}, the function $f(t,u)$ reads:
$$  
f(t,u) = (t-u)^d
$$   
where the parameter $d$ satisfies $ -1/\alpha < d < 1-1/\alpha$. The LFSM is thus
a self-similar process with stationary increments. Further the exponent
$H$ defined earlier is here $=d + 1/ \alpha$ and $\sigma^{\alpha}_t =
\sigma^{\alpha}_1 |t|^{\alpha H}$. Thus $x(t)$ can be seen as an
integral or derivative of fractional order, of the noise $\eta$. 

\begin{figure}
\centerline{
           \psfig{figure=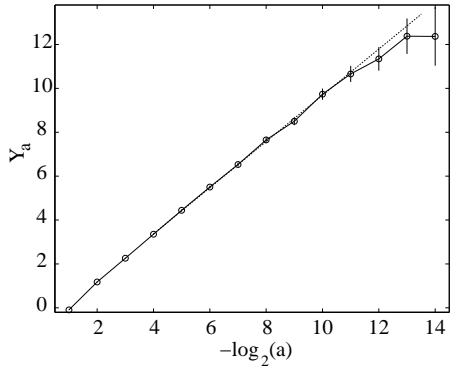,width=6cm,angle=0}}

            \vspace*{0.1 cm}
\caption{
Log-log plot of the amplitude of the wavelet component versus scale factor
using a Daubechies3 wavelet on time series produced by a LREM [6]. 
It exhibits a quasi perfect power-law revealing self-affinity.
The slope is $H+1/2$ where $H$ is 
the self-affinity parameter.}
\label{figloglog}  
\end{figure}

In order to have a prescription for constructing the 
LFSM given the activity plot of the extremal models 
that we studied, it is important to determine carefully 
the $H$ exponent in these models. We have done this by performing
first a wavelet transform \cite{m97}, which consists of
computing wavelet coefficients $d_x(a,k)=\langle x,
\psi_{a,k} \rangle$ where $\psi_{a,k}(t)=
\sqrt{a}\psi_0[a(t-k)]$ is a collection of dilated and translated
templates of the mother wavelet $\psi_0(t)$. 
The wavelet transform is a
relevant tool  \cite{aftv99} to analyse self-similarity because 
the $d_x(a,k)$ of a
self-similar process with stationary increments i)
reproduce exactly the self-similarity, ii) form stationary sequences,
iii) are quasi independent statistically one from another.  
For the LFSM, it has been shown \cite{adf99} that $H$ can be estimated by
performing a linear fit in a $(\log_2(a),Y_a)$ plot, where $Y_a = \langle 
\log_2|d_x(a,k)| \rangle_k $.  
Confidence interval for the estimate of $H$ can be 
theoretically derived \cite{adf99}.
Figure \ref{figloglog} illustrates this estimation procedure on
data produced by an LREM using $b=1$ \cite{Tang}. Only one such
example is shown, but many other trials were performed using other
values of $b$, or other models such as the Bak-Sneppen model.
All the models produced power-laws of similar quality, thus proving that
the activity map of these models can be modelled as an LFSM.

\begin{figure}
\centerline{
           \psfig{figure=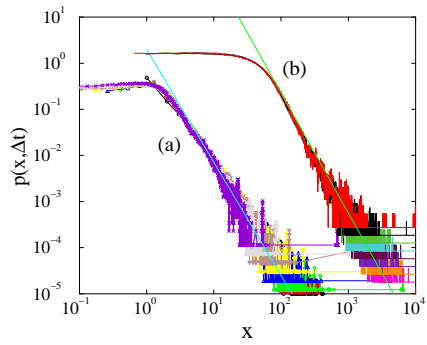,width=6cm,angle=0}}

            \vspace*{0.1 cm}
\caption{
Data collapse of eight different probability distributions 
$p(x,\Delta t)$ with  time intervals $\Delta t$ 
ranging from 1 to 128 (a) for the 
extremal model of \protect \cite{Tang}(L=16384). 
$\alpha = 1.45$ and the  
best data collapse is obtained for $z=1.55$. 
Idem for a LFSM (b) with $\alpha = 1.5$ and 
$H = 0.562$. The best collapse is obtained with $z=1/H$. }
\label{figloglogfur}
\end{figure}

We now try to compare the exponents $\alpha$ and $z$ predicted by the
scaling form (\ref{eq:furub}) obtained for an extremal model (such as
the LREM) and its  LFSM description. 
From 
the definition of the LFSM given by
Eq. (\ref{eq:pLFSM}) and using the fact that
$\sigma^{\alpha}_t = \sigma^{\alpha}_1 |t|^{\alpha H}$, 
it is easy to show that the probability density function
for the LFSM satisfies the scaling form (\ref{eq:furub}) 
with
$z=1/H$. This is also very well verified on numerical simulations of 
LFSMs (fig.\ref{figloglogfur}.b)
and gives an estimate of $z$ with a precision of $\pm 10^{-2}$ 
for a signal of $10^5$ time steps. To compare,
this analysis has also been applied 
on the LREM
(see fig.\ref{figloglogfur}.a) allowing estimates 
on $z$ with a precision  of $\pm 0.1$. The two estimates of $z$ give
us the same result.

To further justify the fact that the LFSM models accurately the activity
map of extremal models, we studied another property, namely the distribution
of return times, which had been characterized on LREM as power-laws, but
never on LFSM.  The first return time is the time $T$ elapsed between two
subsequent activities at a given site.  It has been shown for LREM, that
$T$ is distributed as a power-law $p(T)\propto T^{-\tau_f}$ up to a maximum
time $T^*$ such that the activity has spead over the entire system,
$T^*\propto L^z$.  For the LFSM, we have already seen that the
activity plot generated 
has a dynamic exponent 
$z \equiv 1/H$. When $H<1$, $z$ is also the fractal dimension 
of the activity plot. 
From this it is easy to show that the
first return time exponent (which is simply a one dimensional cut
of a fractal set) should be
\be
\label{eq:H}
\tau_{f}=2-H \equiv 2-1/z.
\ee

This is a relation already known to numerically hold for extremal models
\cite{Maslov}. Hence we see that the two are again identical in this
regard. We have also checked this numerically for different extremal
models and their corresponding LFSM's. It is important to note that,
when $H>1$, $\tau_{FIRST}=1$ for all $H$. 
Thus our identification of the LFSM nature of the activity is indeed
justified, and allows us to access a novel 
property of the LFSM which has never been reported so far 
in the literature.

We now turn to some of the interesting consequences of having an
analytical expression (such as Eq. \ref{eq:pLFSM}) for the space-time
plot of the activity. Consider the two-point
function $P(l_1,l_2)$: the probability of having a jump $l_1$ and
$l_2$ at two consecutive instants. Using Eq. (\ref{eq:pLFSM}) we find that 
the expression of this
function in Fourier space is just $ \sim 
\int \exp(\imath k(l_1+l_2)) \, \exp(-|2|^{\alpha H} |k|^{\alpha}) dk \, . 
$ In real space this is just the function $1/(l_1+l_2)^{\alpha+1}$
asymptotically. We can use this to compute the conditional probability 
$P(l_2/l_1)$ --- the probability of having a jump of length $l_2$ at
time $t=2$ given that there was a jump $l_1$ at time $t=1$ --- using
the well known relation $ P(l_2/l_1) = P(l_1,l_2)/P(l_1)$. We see that
this is just $ \sim 1/(1+l_2/l_1)^{\alpha+1}$ thus theoretically 
confirming the numerical scaling obtained in \cite{Supriya}. 
We can now generalise to the full
$n$-point function $P(l_1,l_2,l_3....l_n)$ (discussed in \cite{Supriya})
of having a jump of length $l_1$ at time $t=1$, a jump of length $l_2$
at time $t=2$ and so on till $t=n$. Systematic expansions can be
obtained for the conditional probabilities just as for the two-point
function and can also be verified numerically on the models. The
theoretical understanding of these correlation functions is hence one
of the achievements of our mapping.

A further very interesting consequence of the definition of the LFSM
is the following equation that we can write down for the front propagation
under this dynamics.
To do this, we define the height of an interface at time $t$ 
at a spatial location
$X$ as simply the accumulated activity there upto time $t$:
\be
\label{eq:hh}
h(X,t)= \Delta h.\int_0^{t} \delta (X- x(t^{\prime})) dt^{\prime} .
\ee

Note that we can use with this definition (\ref{eq:hh}) 
to perform the 
scale transformation commonly used for self-affine surfaces:
$ X \rightarrow bX \, ,\, t \rightarrow b^{z}t $ and 
$ h \rightarrow b^{\chi}h$, where $\chi$ is the so called 
{\it roughness exponent} for the height. 
A power counting on both
sides of this equation gives the relation $ \chi = z -1$, 
known to hold for extremal
dynamics ~\cite{Tang,Maslov}.

Using the definition of the LFSM (\ref{eq:lf}) and 
the following definition of the fractional derivative \cite{Kol}:
\begin{eqnarray}
\label{eq:defrac}
\frac{\partial^{\beta} \psi}{\partial x^{\beta}} (X-x(t)) &\equiv&
\nonumber \\ 
\frac{1}{\Gamma (1-\beta)} 
\lim_{\delta \rightarrow 0} &\displaystyle \frac{d}{d\delta}&
\int^{x(t) +\delta}_{x(t)} 
\frac{\psi(u)-\psi (x(t))}{(x(t)+\delta-u)^{\beta}}
du .
\end{eqnarray}
we obtain the following equation for the height after tedious calculations (the
details of which we leave for a longer paper \cite{LP}):
\begin{eqnarray}
\label{eq:granfin}
\frac{\partial^{\alpha H} h(X,t)}{\partial t^{\alpha H}} &=&
\frac{\Gamma(\alpha H +1)}{\Gamma (\alpha +1)}
\frac{\partial^{\alpha}}{\partial X^{\alpha}} h(X,t)
\nonumber \\
+
\frac{c}{\Gamma(2-\alpha H)} \int^{t}\,{du} . &\eta(u)& . \int_{u}^{t} ~du' 
\frac{\partial^{2} h(X,u')} {\partial X\partial {u'}}  |u'-u|^{d-1}
\end{eqnarray}
In deriving the above, we have basically used some of the techniques
developed for derivatives of fractional order such as the 
fractional Taylor expansion \cite{Kol} and 
fractional integration by parts. $c$ is a constant with $c \sim
(\delta t)^{1-\alpha H}$ taking into account 
the right dimensionality of the noise.

We have thus changed a problem with {\em quenched disorder} to one with
a multiplicative {\em annealed noise}, $\eta(t)$, with our approach. 
There are several points worth
mentioning in this regard. Using the fact that $p(X,t)$ is
proportional to the number of times the site $X$ has been visited at
time $t$ (for different realisations of the process), that
is $p(X,t)\sim \langle {\partial h(X,t)}/{\partial t} \rangle$,  
we can also conclude from Eq. (\ref{eq:granfin}) that an effective 
Fokker Planck equation for the activity
should be one of fractional order in space and time: 
\be
\frac{\partial^{\alpha H} p(X,t)}{\partial t^{\alpha H}} \sim
\frac{\partial^{\alpha}}{\partial X^{\alpha} } p(X,t)  .
\ee
We note that the expression Eq. (\ref{eq:pLFSM}) is consistent with the
above equation in the long-time limit \cite{Klafteretal}.
For an uncorrelated process the above equation reduces to the
one studied in \cite{fogedby}. An
equation such as the above has also been derived for a process
with long-ranged jumps and a power-law waiting time between jumps
\cite{Barkai}. It would be interesting to see the connection between
the above process and Eq. (\ref{eq:lf}) more microscopically.

\begin{figure}
\centerline{
           \psfig{figure=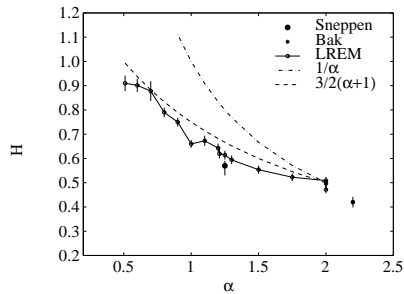,width=6cm,angle=0}}

            \vspace*{0.1 cm}
\caption{
This diagram represents the (wavelet-based) 
estimated $H$s (and their confidence intervals) 
for various $\alpha$ and various extremal models. The dot-dash curve
$1/\alpha$ is the estimate for an uncorrelated levy-flight. The
function $3/2(\alpha +1)$ is the numerically found best estimate for
$z$ in the LREM. The two other extremal models studied however do not
seem to obey this relation. }
\label{figalphah}
\end{figure}

There are also several other interesting points to investigate. 
As we have proved
above, any extremal model can be modelled by a tangent process, which
we call the LFSM, for a suitable value of $H$ and $\alpha$.
In Fig. \ref{figalphah} we
have indicated the above parameter values for all the extremal
models mentioned in this letter. It is interesting to note that for
the models studied in \cite{Tang} 
a further relation exists between $H$ and $\alpha$; $ H=3/2 (\alpha
+1)$ \cite{Tang} as indicated in Fig. 3. 
The two other extremal models studied (The Bak Sneppen model and the
Sneppen Interface model) however do not seem
to obey this relation (Fig. 3). 
It would be interesting to understand why the above relation exists
for the LREM.

In summary we have proposed the LFSM as a stochastic model for
extremal processes. We argue that the 
self-affinity and
$\alpha$-stability of this process, together with the stationarity
of its increments makes it an accurate description
of extremal dynamics. We demonstrate this by first determining the value $H$ 
and $\alpha$ for all the extremal models listed above
by using a  wavelet analysis of the activity.
Subsequently we show that a LFSM process generated using these
values shows the same scaling predicted by Eq. \ref{eq:furub}
because of a simple relation between
the dynamical exponent $z$ and the exponent $H$. In order to test such
a correspondence further we also investigate a property which has
never been studied in the LFSM, namely the time interval distribution
between recurrences of activity at a particular site.
Further the LFSM is amenable to an analytical treatment 
much more easily than 
extremal dynamics. We show how the scaling form 
(\ref{eq:furub}) is a simple consequence of the definition of the process.
We comment on how the process allows us to understand the full $n$-point
probability distribution for the increments. Finally we conclude by
proposing a fractional partial differential equation 
for a front subjected to extremal dynamics.



\newpage

\end{document}